\documentstyle[aps]{revtex}

\newcommand{\DM}{\vec D_{i,j}}
\newcommand{\jj}{$J_2$/$J_1$}
\newcommand{\jnjnn}{$J_1-J_2$\ }
\newcommand{\be}{\begin{equation}}     
\newcommand{\ee}{\end{equation}}
\newcommand{\bdm}{\begin{displaymath}}     
\newcommand{\edm}{\end{displaymath}}
\newcommand{\bea}{\begin{eqnarray}}     
\newcommand{\eea}{\end{eqnarray}}
\newcommand{\beas}{\begin{eqnarray*}}     
\newcommand{\eeas}{\end{eqnarray*}}
\newcommand{\bite}{\begin{itemize}}
\newcommand{\enite}{\end{itemize}}

\begin{document}

\title{The \jnjnn antiferromagnet on the square lattice with
Dzyaloshinskii-Moriya interaction : An exact diagonalization
study}

\author{Andreas Voigt and Johannes Richter}

\address{Institut f\"ur Theoretische Physik,
Otto-von-Guericke-Universit\"at Magdeburg\\ Postfach 4120, D-39016
Magdeburg, Germany}
\date{Received \today}
\maketitle

\begin{abstract}
We examine the influence of an anisotropic interaction term of
Dzyalo\-shinskii-Moriya (DM) type on the groundstate ordering of the
\jnjnn  spin-$1 \over 2$-Heisenberg antiferromagnet on the square
lattice. For the DM term we consider several symmetries corresponding to
different crystal structures. For the pure \jnjnn model there are strong
indications for a quantum spin liquid in the region of $0.4
{\scriptstyle \stackrel{<}{\sim}} J_2/J_1 {\scriptstyle
\stackrel{<}{\sim}} 0.65$. We find that a DM interaction influences the
breakdown of the conventional antiferromagnetic order by i) shifting the
spin liquid region, ii) changing the isotropic character of the
groundstate towards anisotropic correlations and iii) creating for
certain symmetries a net ferromagnetic moment. \end{abstract}



\section{Introduction}
The two-dimensional quantum spin Heisenberg antiferromagnet (AFM)
has attracted a considerable interest in connection with the magnetic
properties of the high-temperature superconductors in recent time
\cite{manou}. The CuO planes being responsible for the superconductivity
show a strong in-plane exchange and only a small off-plane exchange
\cite{shirane}. Therefore in the insulating phase the interacting Cu
spins should be well described by a two-dimensional spin-$1 \over
2$-Heisenberg antiferromagnet. There are several arguments
\cite{inui,annett,ihle,schmidt} that additional to the nearest neighbour
exchange $J_1$ a frustrating diagonal $J_2$ bond is relevant. The groundstate
properties of this so-called \jnjnn model are widely discussed in
the last years, mainly with respect to a possible breakdown of the
magnetic long-range order (LRO) due to the combined influence of quantum
fluctuations and frustration (see e.g.\cite{chandra}-\cite{richter95}).
One finds evidence for a finite parameter region  around \jj $\sim 0.5$
where quantum fluctuations open a window of a spin liquid phase.
Moreover, there are indications of enhanced exotic order parameters
(spin-Peierls, chiral)
\cite{dagotto89,richter91,richter93,singh90,kawara91}.

However, there are also indications for additional anisotropic terms in
the Hamiltonian which could explain the experimentally observed weak
ferromagnetism for instance in La$_2$CuO$_4$ \cite{thio,cheong}. In
general, a small ferromagnetic moment in antiferromagnets may appear in
materials with low crystal symmetry. This tilting of the spins can be
described by adding the so-called anisotropic Dzyaloshinskii-Moriya (DM)
interaction to the isotropic Heisenberg model. Already 1957
Dzyaloshinskii formulated a phenomenological theory of these facts
\cite{dzialo}. Three years later Moriya developed the microscopic theory
of the weak ferromagnetism \cite{moriya}. The occuring additional
interaction term in the Hamiltonian is proportional to the DM vector
$\vec D$.  In some recent publications
\cite{coffey90,coffey91,bone,koshi,feldkemper} one has examined the
origin and the structure of $\vec D$. In this paper we discuss the
influence of DM terms on the groundstate properties of the \jnjnn model
on the square lattice.


\section{Anisotropic spin interaction}

Let us start with a general Hamiltonian, describing quadratic spin-spin
interaction:

\be
\label{hamal1}
\hat H = \sum_{i,j=1}^{N} {\hspace{0.2cm}} \vec S_i^{{ }T} {\hspace{0.3cm}}
\hat J_{i,j} {\hspace{0.3cm}} \vec S _j
\ee
with $\hat J_{i,j}$ being a 3x3 matrix of interaction, which can be
written as
\bea
\label{jij}
\hat J
= {1 \over 3} (Sp \hat J) \hat 1 + \hat A + \hat M \hspace{.2cm} ,
\eea
where $\hat 1$ is the unit matrix, $\hat A$ is antisymmetric and
$\hat M$ is symmetric and traceless. With (\ref{jij}) equation
(\ref{hamal1}) reads:
\bea
\label{hamjdm}
\hat H = \sum_{i,j=1}^N \bigl \{ {1 \over 3} (Sp \hat
J_{i,j}) \  \vec S_i  \vec S_j  + \vec D_{i,j} ( \vec S_i \times \vec
S_j )
+ \vec S_i^T \hat M_{i,j} \vec S_j \bigr \} \hspace{.2cm} ,
\eea
where the first term is the isotropic Heisenberg interaction, the second
one - the Dzyaloshinskii-Moriya interaction and the last one - the
anisotropic pseudo-dipole interaction. The vector $\vec D_{i,j}$
contains the three independent components of $\hat A_{i,j}$. In general
the occurrence of $\vec D_{i,j}$ requires low crystal symmetries. The
origin of the anisotropic interaction $\DM$ and $\hat M_{i,j}$ is the
spin-orbit coupling. In the last few years there is some effort
concerning the question of including the term with $\hat M_{i,j}$ in the
calculations \cite{shekt}, but in what follows we neglect $\hat
M_{i,j}$, because it is only of second order in the spin-orbit coupling
constant $\lambda$ whereas $\DM$ is of first order in $\lambda$.

The weak ferromagnetic moment in the predominantly antiferromagnetic
ordered CuO planes in La$_2$CuO$_4$ can be interpreted with a small spin
canting. This spin canting could be described via a DM interaction term.
In \cite{coffey90} a general form of the DM vector is introduced and it
is shown, that only a vector $\vec D_{i,j}$ which varies from bond to
bond corresponds to the crystal structure and is able to describe the
observed weak ferromagnetism. In \cite{coffey90,coffey91} Coffey and
coworkers consider different crystal symmetries and get $\DM$ for the
whole lattice by requiring for the vector $\DM$ that the energy of any
configuration of spins is invariant under the symmetry transformations
of the crystal structure. Once  $\DM$ is fixed on a single bond, the
symmetries determine $\DM$ on the entire lattice.

Following \cite{coffey90,coffey91} we consider in this paper two
different symmetries for $\DM$ which correspond to the orthorhombic and
tetragonal phases of La$_2$CuO$_4$ \cite{longo,grande,axe}. The
arrangements of the atoms in a CuO-plane are presented in figure 1.
For the orthorhombic arrangement (figure 1a) one finds the following $\DM$:

\begin{center}
\begin{minipage}{8cm}
$\bullet$ $\vec D_{A,B}=(d_1,d_2,0)$ \\
$\bullet$ $\vec D_{A,C}=(-d_2,-d_1,0)$ \\
$\bullet$ $\vec D_{B,D}=- \vec D_{A,C}=(d_2,d_1,0)$\\
$\bullet$ $\vec D_{C,D}=- \vec D_{A,B}=(-d_1,-d_2,0)$\\
$\bullet$ $\vec D_{C,E}=- \vec D_{A,C}=(d_2,d_1,0)$    \\
\end{minipage}
\end{center}
For the tetragonal arrangement (figure 1b) the $\DM$ is as follows:
\nopagebreak[4]
\begin{center}
\begin{minipage}{8cm}
$\bullet$ $\vec D_{A,B}=(0,d_1,0)$                       \\
$\bullet$ $\vec D_{A,C}=(0,d_2,0)$                      \\
$\bullet$ $\vec D_{B,D}=- \vec D_{A,C}=(0,-d_2,0)$        \\
$\bullet$ $\vec D_{C,D}=- \vec D_{A,B}=(0,-d_1,0)$        \\
$\bullet$ $\vec D_{C,E}=- \vec D_{A,C}=(0,-d_2,0)$        \\
\end{minipage}
\end{center}
Furthermore we will distinguish between equal and different signs of the
parameters $d_1$ and $d_2$. Following the arguments of \cite{coffey90}
we will restrict our considerations to $d_1$ and $d_2$ of equal
strength. Let us define four different cases for the parameters:
orthorhombic symmetry with either $d_1 =+d_2$ or $d_1 =-d_2$ and
tetragonal symmetry with either $d_1 =+d_2$ or $d_1 =-d_2$.

First let us consider the classical Hamiltonian
\bea
\label{H_DM_clas}
\hat H_{DM}= \sum_{i,j=1}^N \vec D_{i,j} ( \vec S_i \otimes \vec S_j )
\eea
with the configurations for $\DM$ as given above. $\vec S_i$ and $\vec
S_j$ are classical 3D vectors and the summation runs over nearest
neighbour bonds. The classical groundstate of (\ref{H_DM_clas}) was
discussed in detail in \cite{coffey90}. Here we briefly summarize some
ground state features which are relevant for the further discussion.
The energy of any nearest neighbour bond $(i,j)$ on the lattice can be
minimized by spin vectors $\vec S_i$ and $\vec S_j$ perpendicular to
each other and perpendicular to the DM vector $\vec D_{ij}$; the
corresponding optimum bond energy is $E_{ij} = - \mid \vec D_{ij} \mid
S^2 $. Because of the special symmetry $\mid d_1 \mid = \mid d_2 \mid$
there is no frustration in the Hamiltonian (\ref{H_DM_clas}) and the
total groundstate configuration can be build by the suitable arrangement
of optimized bonds. In table 1 we present some important groundstate
features for the considered four different cases of $\DM$.

\begin{center}
\begin{minipage}[b]{10cm}
\begin{itemize}
\item[] \begin{tabular}{@{}c|llcc}  case &
crystal & DM & net & spin \\ \vspace{-.5cm} \\
& structure & vector & FM & configuration \\ \hline
Ia & orthorhombic & $d_1 =+d_2=d$ & no  & 4 SL \\
Ib & orthorhombic & $d_1=-d_2=d$ & yes & 2 SL \\
IIa & tetragonal & $d_1 =+d_2=d$  & yes & 2 SL \\
IIb & tetragonal & $d_1=-d_2=d$  & no & 4 SL \\
\end{tabular}
\item[] Table 1: The four considered cases with some of its groundstate
features of the classical DM Hamiltonian $\hat H_{DM}$
\end{itemize}
\end{minipage}
\end{center}

Here FM means ferromagnetic moment and 2 SL or 4 SL means 2 or 4
sublattices, respectively. As an example the groundstates for the cases
Ia and Ib are illustrated in more detail in figure 2.

In the next section we turn to the quantum system.


\section{The Model}

We start with the so called \jnjnn model on the square lattice:

\bea
\label{hamj1j2} 
\hat H_{J_1,J_2} = \sum_{i=1}^N J_1 \left( \vec S _i \vec S _{i+ \vec x}
+ \vec S _i \vec S _{i+ \vec y} \right) + J_2 \left( \vec S _i
\vec S_{i+ \vec x + \vec y} + \vec S _i \vec S _{i - \vec x + \vec y}
\right).
\eea
$\vec S_i$ denotes the spin-${1 \over 2}$ operator on site i and
$\vec x$ and $\vec y$ are the unit lattice vectors in x- and
y-direction.

An intensive investigation over the last years (see
e.g.\cite{chandra}--\cite{richter95}) suggests the following phase
diagram: For small values of $J_2 /J_1$ the system shows
AFM-LRO. At some critical value $J_2^{crit}/J_1$ of about $ \simeq 0.4$
the conventional collinear AFM-LRO breaks down. Then a region of a
spin-liquid state could be realized for $0.4 \le J_2/J_1 \le 0.65$. For
values of $J_2/J_1 \ge 0.65$ an AFM-LRO arises within the initial
sublattices (4-sublattice AFM).

Now we add the Hamiltonian of the DM-interaction

\bea
\label{htot}
\hat H = \hat H_{J_1,J_2} + \hat H_{DM}
\eea
\bea
\label{hdm1}
\hat H_{DM} = \sum_{i=1}^N  \left[ \vec D_{i,i+ \vec x} (\vec S _i \times
\vec S _{i + \vec x}) + \vec D_{i,i+ \vec y} (\vec S _i \times
\vec S _{i + \vec y}) \right].
\eea
Using spin-flip operators we can rewrite the DM term
\bea
\label{dmterm}
\vec D_{i,j} (\vec S_i \times \vec S_j) & = & {1 \over 2} \bigl \{ i
D_{ij}^x \bigl [ S_i^- S_j^z - S_i^+ S_j^z - S_i^z S_j^- + S_i^z S_j^+
\bigl  ]  \\ \nonumber
& - &  D_{ij}^y \bigl [ S_i^- S_j^z - S_i^z S_j^+ - S_i^z S_j^- + S_i^+ S_j^z
\bigl ]   \\ \nonumber
& + & i D_{ij}^z \bigl [ S_i^+S_j^- - S_i^- S_j^+ \bigl ] \bigl \}
\eea
The Hamiltonian $H_{DM}$ contains terms of the form $S_i^{\pm} S_j^z$
which do not commute with the z-component of the total spin $\vec S$.
Hence we have to take {\bf all} $2^N$--Ising states for the construction
of the wave function. We use a modified Lanczos procedure to calculate
the groundstate of the system.


\section{Numerical results}

Based on the properties of the pure \jnjnn model we discuss several
order parameters, cf.\cite{dagotto89,schulz,richter93}, which describe
the relevant magnetic properties of the model.

For dominating $J_1$ the corresponding antiferromagnetic LRO parameters
are the components of the square of the staggered magnetization:

\bea
\label{ms2}
\left( M_s^\gamma \right)^2  = \left< { \left[ \frac{1}{N}  \sum_{i=1}^{N}
            { S_i^{\gamma} \tau_i } \right]^2} \right>
            \hspace{6pt} ; \hspace{6pt}
            \gamma = x,y,z
            \hspace{6pt} ; \hspace{6pt}
             \tau_i =  \left\{ \begin{array}{r@{\quad\mbox{for}\quad}l}
             1 & i \epsilon A \\
            -1 & i \epsilon B  \end{array} \right.
\eea
where $S_i^{\gamma}$ is the $\gamma$-component of the spin on site i,
$\tau_i$ is the corresponding staggered factor and A and B are the two
sublattices. Obvious these parameters describe the ordinary 2-sublattice
antiferromagnetic ordering.

For dominating $J_2$ the relevant order parameters are

\bea
\label{ms2alpha}
\left( M_{s,\alpha}^\gamma  \right)^2  = \left< { \left[ \frac{1}{N/2}
            \sum_{ i_{\alpha}=1}^{N/2}
            { S_{i}^{\gamma} \tau_{i,\alpha} } \right]^2} \right>
            \hspace{5pt} ;\hspace{5pt} \gamma = x,y,z
            \hspace{6pt} ; \hspace{5pt}
            \tau_{i,\alpha} =  \pm 1 \hspace{5pt} ; \hspace{5pt}
            \alpha=A,B  \hspace{5pt} .
\eea
The sum runs over the sites of the sublattice $\alpha$. These order
parameters describe antiferromagnetic ordering {\bf within} the two
initial sublattices A and B, when the whole system forms a 4-sublattice
AFM. For the rotational invariant model without DM interaction the x,y,z
components of the respective parameters are identical, but the DM
interaction breaks this rotational symmetry.

For $\vec D \neq 0$ a weak ferromagnetism may occur. The relevant
parameters for describing this weak ferromagnetism are the components of
the square of the magnetization,

\bea
\label{m2}
\left( M^\gamma \right)^2  = \left< { \left[ \frac{1}{N}  \sum_{i=1}^{N}
            { S_i^{\gamma} } \right]^2} \right> \hspace{6pt} ;\hspace{6pt}
            \gamma = x,y,z \hspace{6pt} .
\eea
Obviously $\sum_\gamma \left( M^\gamma \right)^2$ is the expectation value
of the square of the total spin $\left< \vec S^2 \right>$.

In the pure model ($\DM = 0$) the Hamiltonian (\ref{htot}) commutes with
$\vec S^2$ and $ S^z$ and the groundstate is isotropic with $\vec S^2
=0$ (and consequently $(M^\gamma)^2 =0$). If we now put on the
anisotropic DM interaction ($\DM \not= 0$) the Hamiltonian does not
commute with $\vec S^2$ and we may expect  $(M^\gamma)^2 \neq 0$.
Koshibae {\it et al} \cite{koshi} pointed out that anisotropic spin
correlations are a result of the interplay between quantum fluctuations
and the anisotropic DM interaction. This can be illustrated as follows:
Consider a classical spin system without DM interaction and a collinear
spin structure. The DM term contains the cross product of the spins so
it vanishes. If now the quantum fluctuations are put on, the DM
interaction favours the direction of spin fluctuations with minimal
energy. Of course this direction depends on the DM vector but
additionally on the classical collinear structure.

We have calculated the groundstate for a square lattice with $N=16$ and
$N=20$ sites for the four different configurations of the DM vector given
in table 1. In the following we focus our interest on the orthorhombic
symmetry, since this symmetry is valid for low doping in
La$_{2-x}$Ba$_x$CuO$_4$ \cite{axe}. Furthermore we present preferably
the data for $N=20$. The corresponding data for $N=16$ are qualitatively
the same. For simplicity we put $J_1=1$ for the rest of the paper.

First we consider the situations of $J_2=0$ and of $J_2=1$. Then the
isotropic model ($\DM =0$) is collinear long-range ordered with either a
2-sublattice structure ($J_2=0$) or a 4-sublattice structure ($J_2=1$).
The influence of the DM interaction on the magnetic order parameters is
shown in figures 3 and 4.

Figure 3 shows the different components of the ferromagnetic moment
$(M^\gamma)^2$ as a function of the DM interaction for $J_2=0$. The
results are as follows: The DM interaction causes a small ferromagnetic
moment ($(M^\gamma)^2 \neq 0$) which increases with $d$. This
moment is anisotropic: ($(M^z)^2 > (M^{x,y})^2$. The ferromagnetic
moment is maximal for case Ib which is in accordance with the
classical situation (see table 1). We have calculated the same parameters
for $J_2=1$. In that case the classical structure is a 4-sublattice AFM,
which does not have a net ferromagnetic moment. Consequently we find in
the quantum case that the ferromagnetic moment is about 10 times smaller
as for $J_2=0$.

In the next two figures 4a and 4b we present the different components of
the AFM-LRO parameters. For $J_2=0$ the 2-sublattice AFM-LRO parameters
$\left( M_s^\gamma \right)^2$ have to be considered and for $J_2=1$ the
4-sublattice AFM-LRO parameters $\left( M_{s,\alpha}^\gamma \right)^2$
are relevant. As discussed above the anisotropy depends on the DM vector
$\DM$ and on the underlying magnetic structure. Correspondingly the
anisotropy is changed going from $J_2=0$ to $J_2=1$ or, alternatively,
going from case Ia to case Ib. Furthermore, in dependence on the
symmetry of the DM vector, we can observe either an enhancement or a
suppression of a certain component of the antiferromagnetic order
parameter. For instance the z-component of the 4-sublattice AFM-LRO
parameter $\left( M_{s,\alpha}^z \right)^2$ is enhanced by the DM
interaction if $\DM$ favours a classical 4-sublattice structure (see
figure 4a) whereas the same order parameter is suppressed if $\DM$
favours a classical 2-sublattice structure (see figure 4b).

The comparison of the magnitude of ferromagnetic and antiferromagnetic
order parameters (figures 3 and 4) shows that the ferromagnetic moment
remains small even if the DM interactions reaches the same strength as
the exchange coupling.

Let us now discuss the antiferromagnetic order parameters in dependence
on the frustration parameter \jj. For this we select from figure 4 in each
case the dominating components $\left(M_{s}^{\gamma} \right)^2$ and
$\left(M_{s,\alpha}^{\gamma} \right)^2$. The magnetic order for $\DM=0$
is determined by the competition between $J_1$ and $J_2$. For dominating
$J_1$ and for dominating $J_2$ the magnetic structure is collinear
antiferromagnetic with two or four sublattices. The transition between
both phases is in the classical case at $J_2 = {1\over2} J_1$ and may be
connected with a spin liquid phase in the quantum case. Since the DM
term also favours a two (case Ib) or a four sublattice structure (case
Ia) we expect that this competition is influenced if we put on the DM
term. The result should be a shift of the transition either to lower or
higher values of $J_2$.

As shown in figure 5 we find indeed the expected shift, it is in
particular strong for case Ia. As discussed above, the critical
$J_2$ for the pure model ($\DM$=0) is about $J_2 \approx 0.4$, which is
too large to be realistic for the cuprate superconductors. We conclude
that a DM interaction of this symmetry is able to support the breakdown
of the antiferromagnetic long-range order due to frustration and to
shift the critical $J_2$ to more realistic values. (Notice, that the
abruptness of the change in the order parameters seen in figure 5 as
well as in figure 6 is due to the well-known level crossing for the
$N=20$ lattice (see e.g. \cite{dagotto89},\cite{richter91})).

Finally we will discuss the region of a possible spin liquid in more
detail. As pointed out in \cite{dagotto89,richter91,richter93,singh90}
there are two interesting candidates for unconventional noncollinear
ordering in this region. One is a vector chiral order parameter
introduced in \cite{richter91,richter93},

\bea
\label{chi}
C^\gamma & = & { \left\langle \frac{1}{3} \left[ \frac{1}{2N}
\sum_{i=1}^N
 { ({\vec C}_{i,i+\vec{x},i+\vec{x}+\vec{y}}^\gamma -
 {\vec C}_{i,i+\vec{x}+\vec{y},i+\vec{y}}^\gamma ) }
  \right]^2\right\rangle} \\
{\vec C}_{ijl}^\gamma & = & 8 \vec e_\gamma \left( ( \vec S_i \times
\vec S_j ) + ( \vec S_j \times \vec S_l )  + ( \vec S_l \times \vec S_i)
\right) .
\eea
This order parameter measures the handiness of a plaquette of three
spins. For the isotropic Heisenberg system this parameter is the same
for all three components x,y and z. But in our case of anisotropic DM
interaction the different components may have different values.

Another candidate for exotic ordering is the dimer (or spin-Peierls)
order parameter,

\bea
\label{dimer}
D = 2 \left [ {1 \over N} \sum_{i=1}^N (-1)^{i_{\vec x}} \vec S_i \vec
S_{i+ {\vec x}}  \right ]^2.
\eea
which measures the long-range phase coherence of spin dimers (singlets
of two spins) along the x (or y) direction.

In figure 6 we present these parameters $C^\gamma$ and $D$, where we
have selected the largest component of $C$. Both order parameters show
in the pure model a characteristic maximum at about $J_2=0.55$. The
influence of the DM interaction is twofold. First the maxima are shifted
in the same way as discussed for figure 5. Second the maxima are
suppressed particularly for $D$. The suppression of the maximum in $D$
can be simply understood by having in mind that the dimer ordering is
connected with a singlet groundstate which is obviously not realized
for $\DM \ne 0$. On the other hand, the effect of the DM interaction on
the chiral order parameter $C^\gamma$ seems to be more complicated. The
DM term creates a spin canting which might under certain circumstances
support the realization of a vector chiral ordering. Indeed, we found
one example ($N=16$, $d_1=+d_2=0.3$), where the chiral order parameter
is enhanced in a small region around $J_2 \approx 0.4$
(figure 6c). Here the DM interaction term together with the frustrating
$J_2$ causes some canted groundstate with enhanced vector chirality.
However, it remains unclear whether this increase of $C^\gamma$ is an
artifact of the particular cluster symmetry of $N=16$.


\section{Summary}

The magnetic properties of the \jnjnn model with anisotropic
Dzyaloshinskii-Moriya (DM) interaction are determined by the competition
between terms favouring a collinear 2-sublattice AFM and terms favouring
a collinear 4-sublattice AFM. The former one is stabilized by the $J_1$
term and the latter by the $J_2$ term. The DM term may favour the
2-sublattice structure as well as the 4-sublattice structure
in dependence on the
symmetry of the DM vector $\DM$ (see table 1 in section 2). As a result
the DM interaction shifts the spin liquid region (which separates the
two collinear antiferromagnetic phases for small and large $J_2$) to
larger or smaller values of $J_2$. Because  the critical $J^{crit}_2
\approx 0.4$ for the breakdown of the 2-sublattice LRO in the pure model
($\DM$=0) is much larger than realistic $J_2$ values for cuprate
superconductors \cite{annett,ihle,schmidt} additional mechanisms are
needed to shift the transition to smaller values of $J_2$. Recently it
has been shown \cite{richter93} that static holes simulating doping can
shift $J^{crit}_2$ in the desired direction. In this paper we find that
a DM term with suitable symmetry acts in the same direction: In case of
orthorhombic symmetry a DM vector with identical x- and y-components
(case 1a in table 1) is appropriate to weaken the 2-sublattice
antiferromagnet and to realize lower $J^{crit}_2$.

Besides of the shift of the transition region the DM term creates an
anisotropy in the spin correlations. Which components are enhanced and
which are suppressed is a result of the interplay between quantum
fluctuations, the particular symmetry of the DM vector as well as the
underlying magnetic structure of the pure model. Hence it occurs that
for the 2-sublattice AFM another component is favoured than for the
4-sublattice AFM.

The third remarkable effect of the DM interaction is the appearance of a
weak ferromagnetic moment. For the considered finite lattices this
moment occurs for all symmetries of $\DM$. In accordance with the
classical picture the orthorhombic structure with different x and y
components of $\DM$ (case Ib in table 1) yields the strongest
ferromagnetic moment and might indicate the existence of weak
ferromagnetism in the thermodynamic limit just for this symmetry of
$\DM$.

Finally we analyzed exotic dimer and vector chiral order parameters
showing a characteristic maximum in the spin liquid region. This
maximum is in general suppressed by the DM term.

Besides of the most important orthorhombic symmetry considered
preferably in the paper we have also calculated the magnetic quantities
for the tetragonal symmetry (cf. figure 1b and table 1). In principle the
same scenario as for the orthorhombic symmetry is valid, i.e. in
dependence on the strength and the symmetry of the DM vector $\DM$ we
may have anisotropy in the spin correlations, a net ferromagnetic moment
and a shift of the spin liquid phase.


\section*{Acknowledgment}

This work has been supported by the Deutsche Forschungsgemeinschaft
(Project No. Ri 615/1-2).



\begin{figure}
\caption{The orthorhombic (a) and tetragonal (b) arrangements of the
CuO-octahedra taken from [24]. The filled circles
are the copper sites which carry the spins, the open circles are oxygen
sites, which are tilted up out of the plane and the crossed circles are
oxygen sites, which are tilted down out of the plane. The square
represents one tilted CuO-octahedra. The points A--E denote certain
copper sites used in the text. The arrows show one particular arrangement
of the DM vector $\DM$ for this lattice (cases Ia and IIa, see table 1.}
\label{Figure 1}
\end{figure}
\begin{figure}
\caption{The classical spin configurations for the Hamiltonian
(\protect{\ref{H_DM_clas}}); (a): case Ia with a 4-sublattice
configuration and without a net ferromagnetic moment, (b): case Ib with
a 2-sublattice configuration and with a net ferromagnetic moment. (All
spins are aligned either along the z-axis or in the xy-plane.)}
\label{Figure 2}
\end{figure}
\begin{figure}
\caption{The x,y and z components of square of the magnetization
$(M^\gamma)^2$ versus $d/J_1$ for case Ia (full line) and case
Ib (dashed line) for $N=20$ and $J_2=0$.}
\label{Figure 3}
\end{figure}
\begin{figure}
\caption{The 2-sublattice AFM-LRO parameters $\left( M_s^\gamma
\right)^2$ for $J_2=0$ (full lines) and the 4-sublattice AFM-LRO parameters
$\left( M_{s,\alpha}^\gamma \right)^2$ for $J_2=1$ (dashed lines)
versus $d/J_1$ for $N=20$; (a): case Ia, (b): case Ib .}
\label{Figure 4}
\end{figure}
\begin{figure}
\caption{The dominant 2-sublattice AFM-LRO parameters $\left( M_s^\gamma
\right)^2$ (full lines) and the dominant 4-sublattice AFM-LRO parameters
$\left( M_{s,\alpha}^\gamma \right)^2$ (dashed lines) versus \jj \ for
$N=20$ and different values of $d$; (a): case Ia with $\left(
M_{s}^{x,y} \right)^2$ and $\left( M_{s,\alpha}^{z} \right)^2$, (b):
case Ib with $\left( M_{s}^{z} \right)^2$ and $\left( M_{s,\alpha}^{x,y}
\right)^2$.}
\label{Figure 5}
\end{figure}
\begin{figure}
\caption{Exotic order parameters versus \jj \ for different $d$: (a) --
The dominant chiral order parameter $C^{x,y}$ for case Ia (solid line)
and $C^z$ for case Ib (dashed line) for $N=20$; (b) -- The dimer order
parameter $D$ for case Ia (solid line) and case Ib (dashed line) for
$N=20$ ; (c) -- The dominant chiral order parameter $C^{x,y}$ for case
Ia (solid line) and $C^z$ for case Ib (dashed line) for $N=16$.}
\label{Figure 6}
\end{figure}

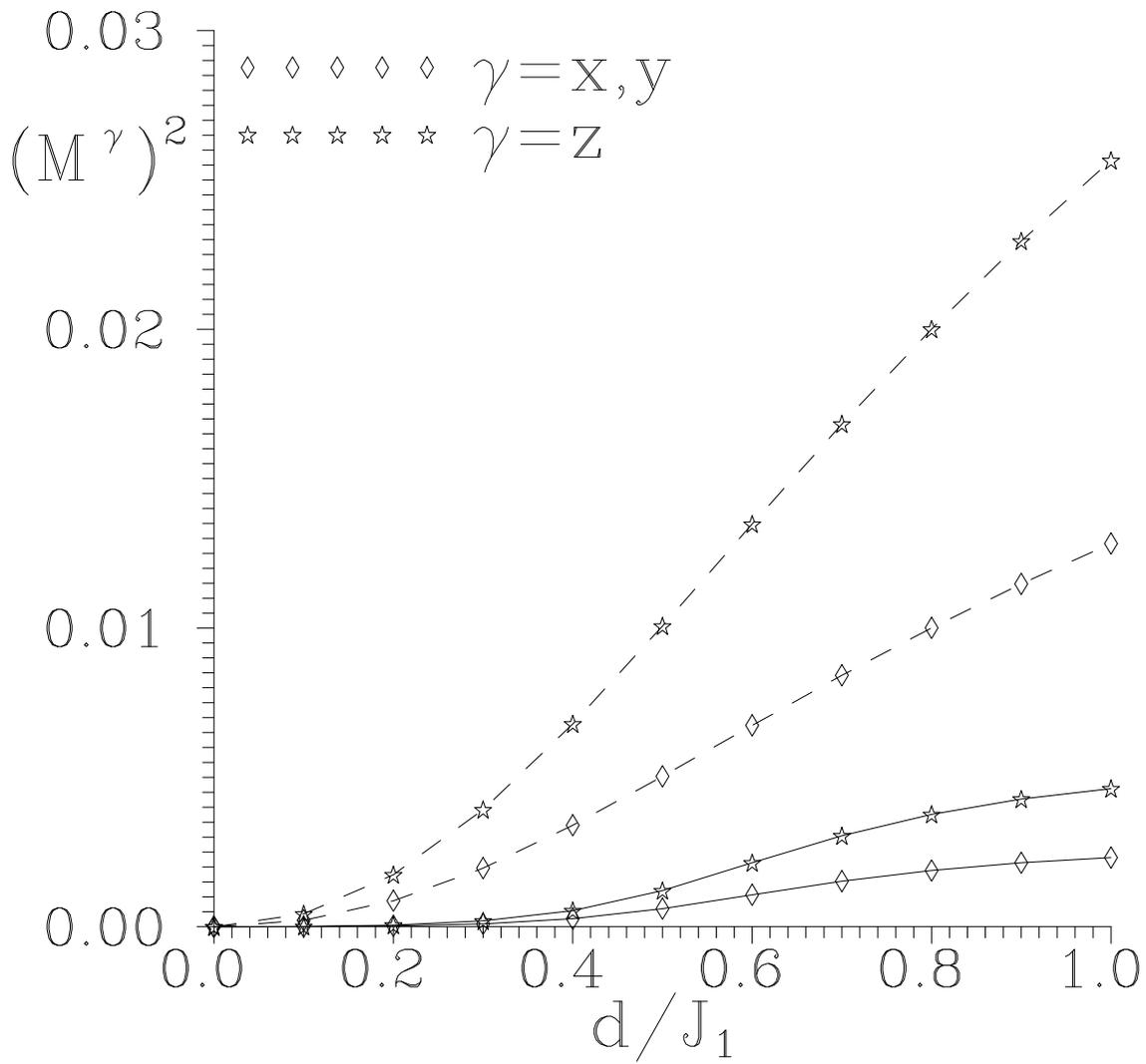
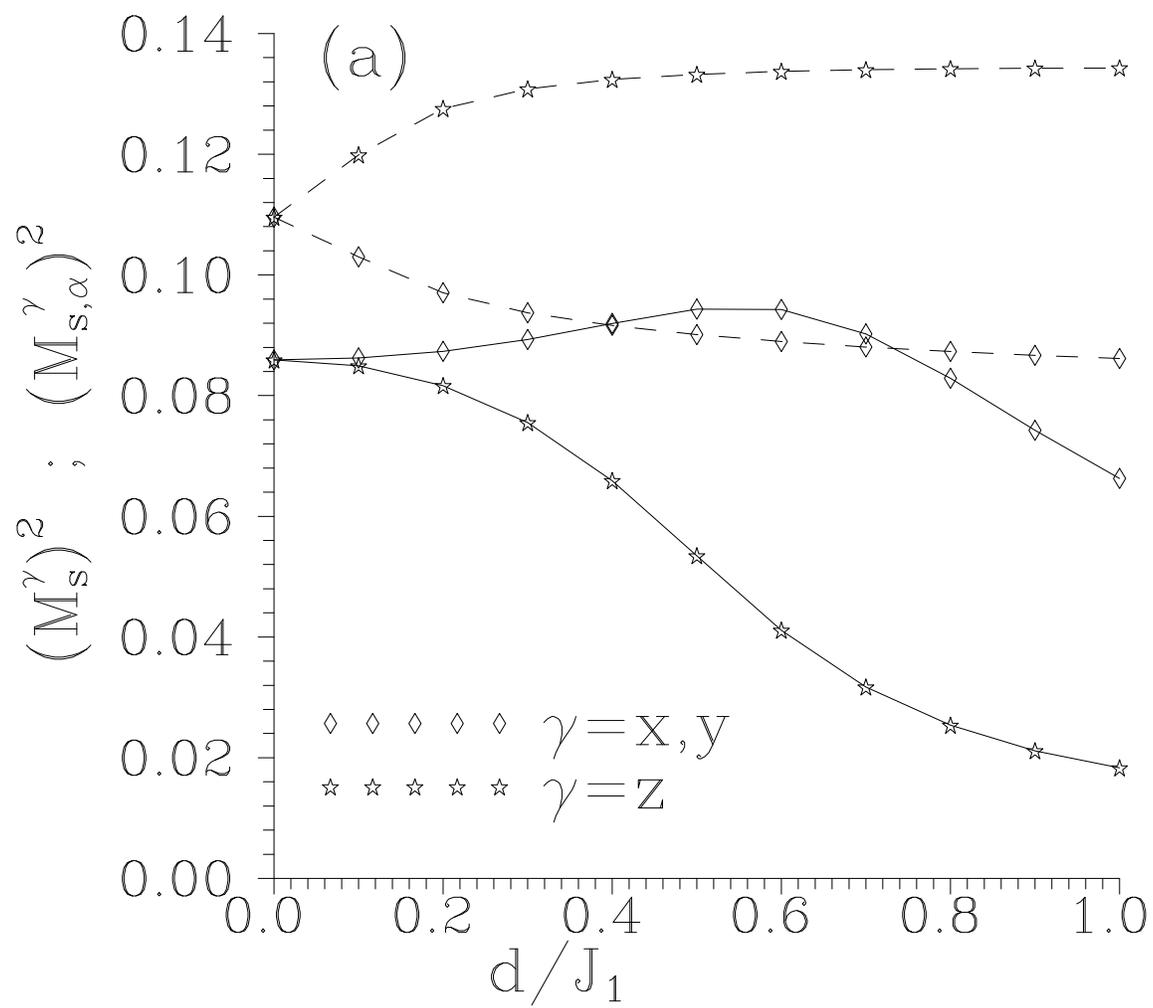
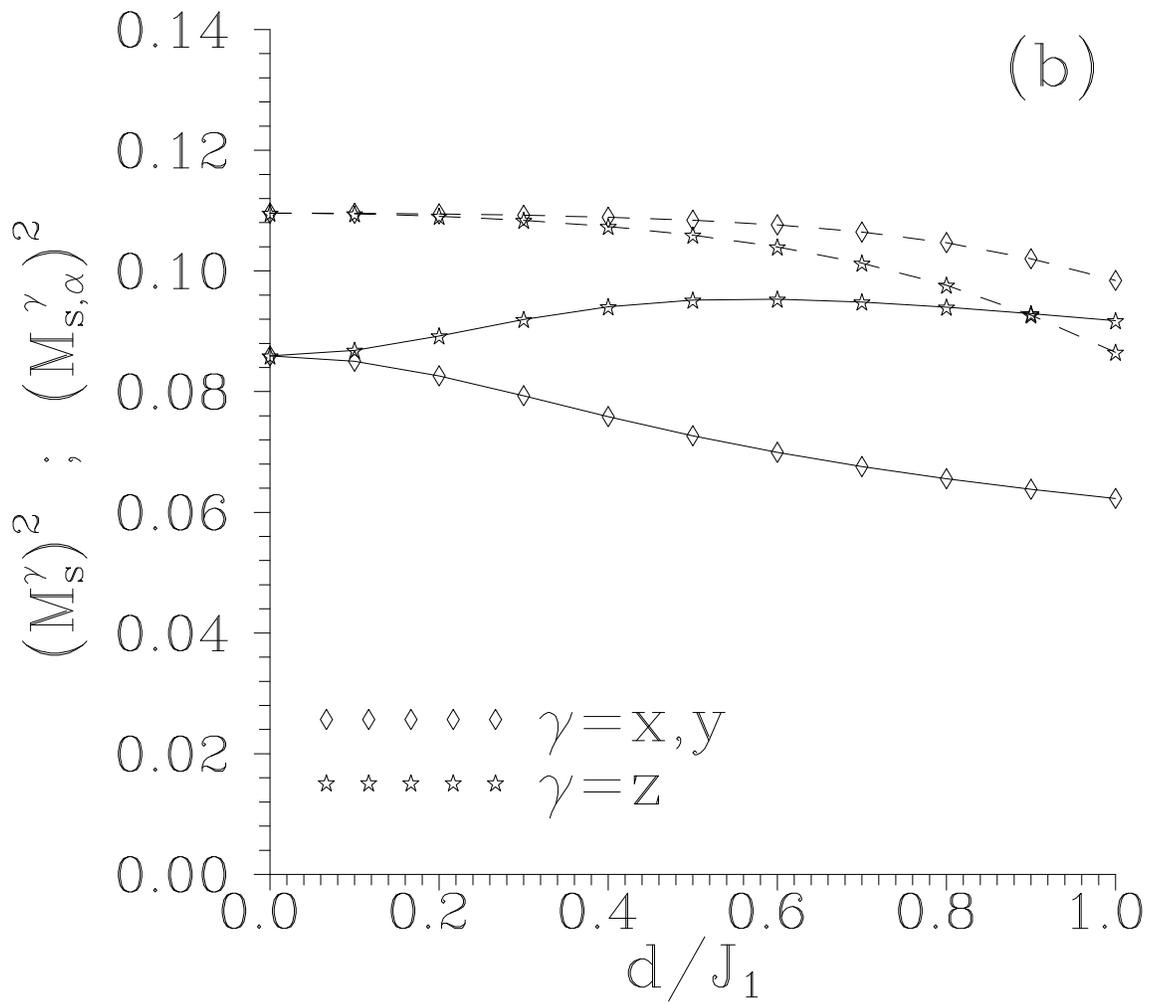
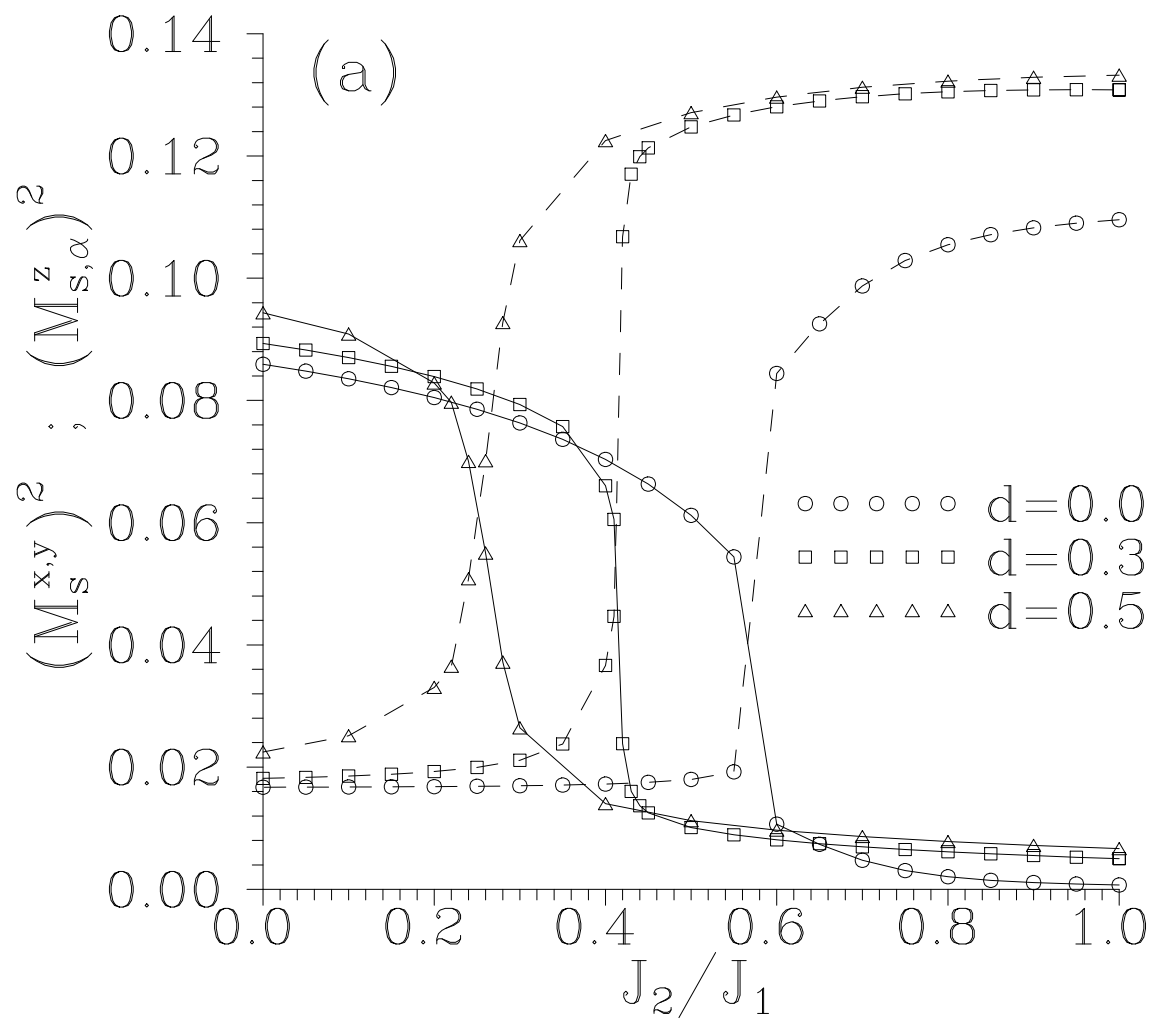
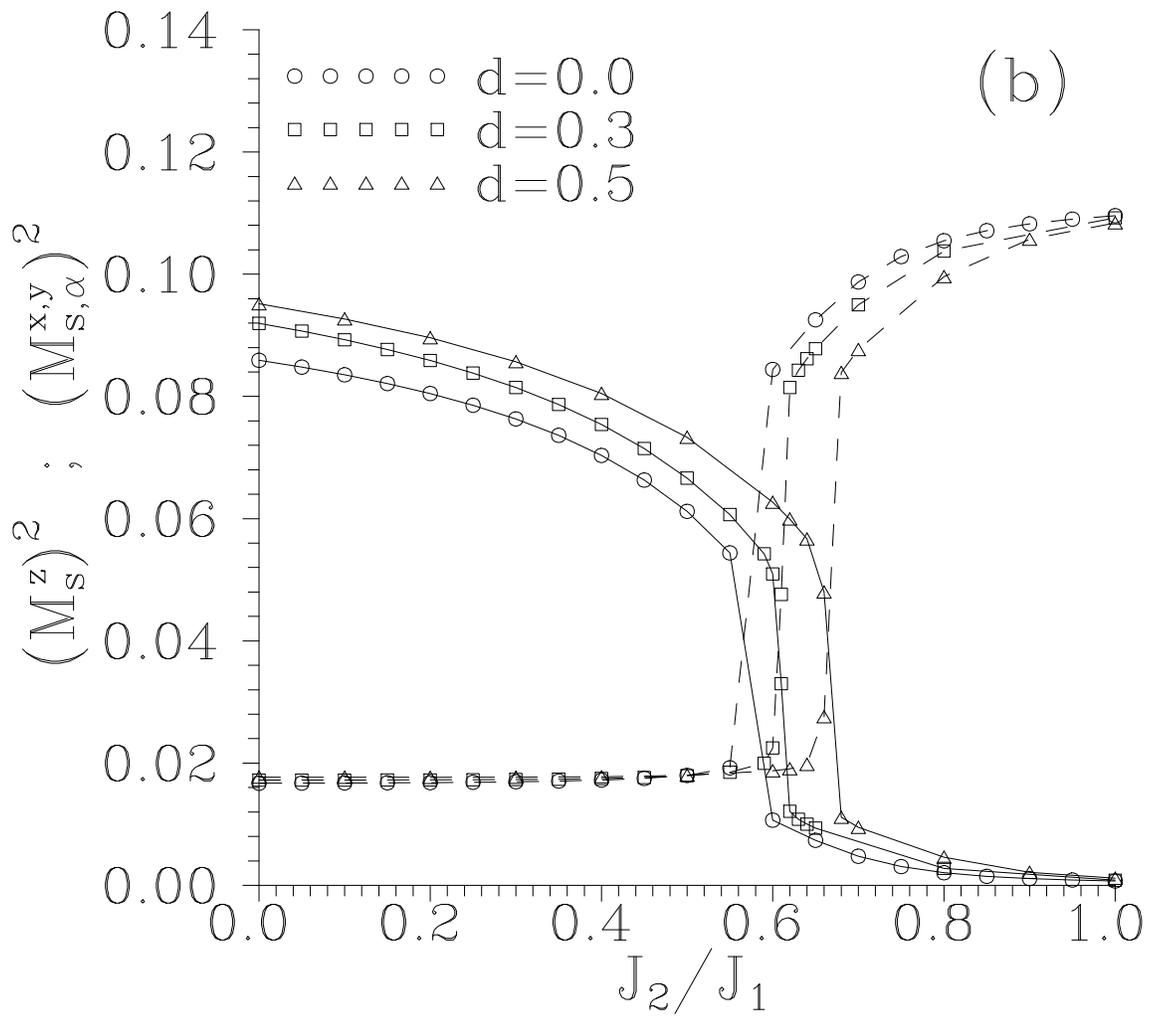
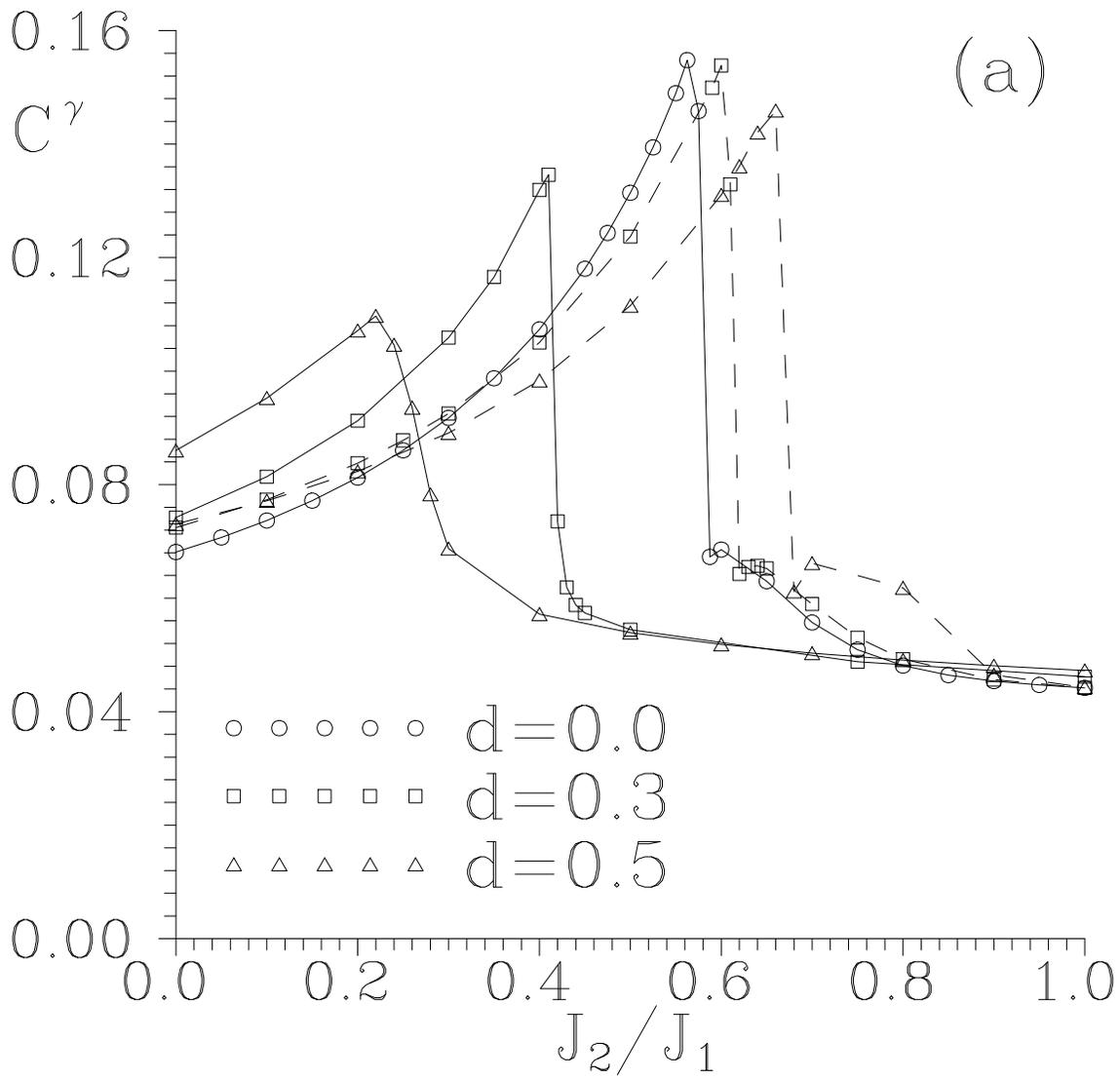
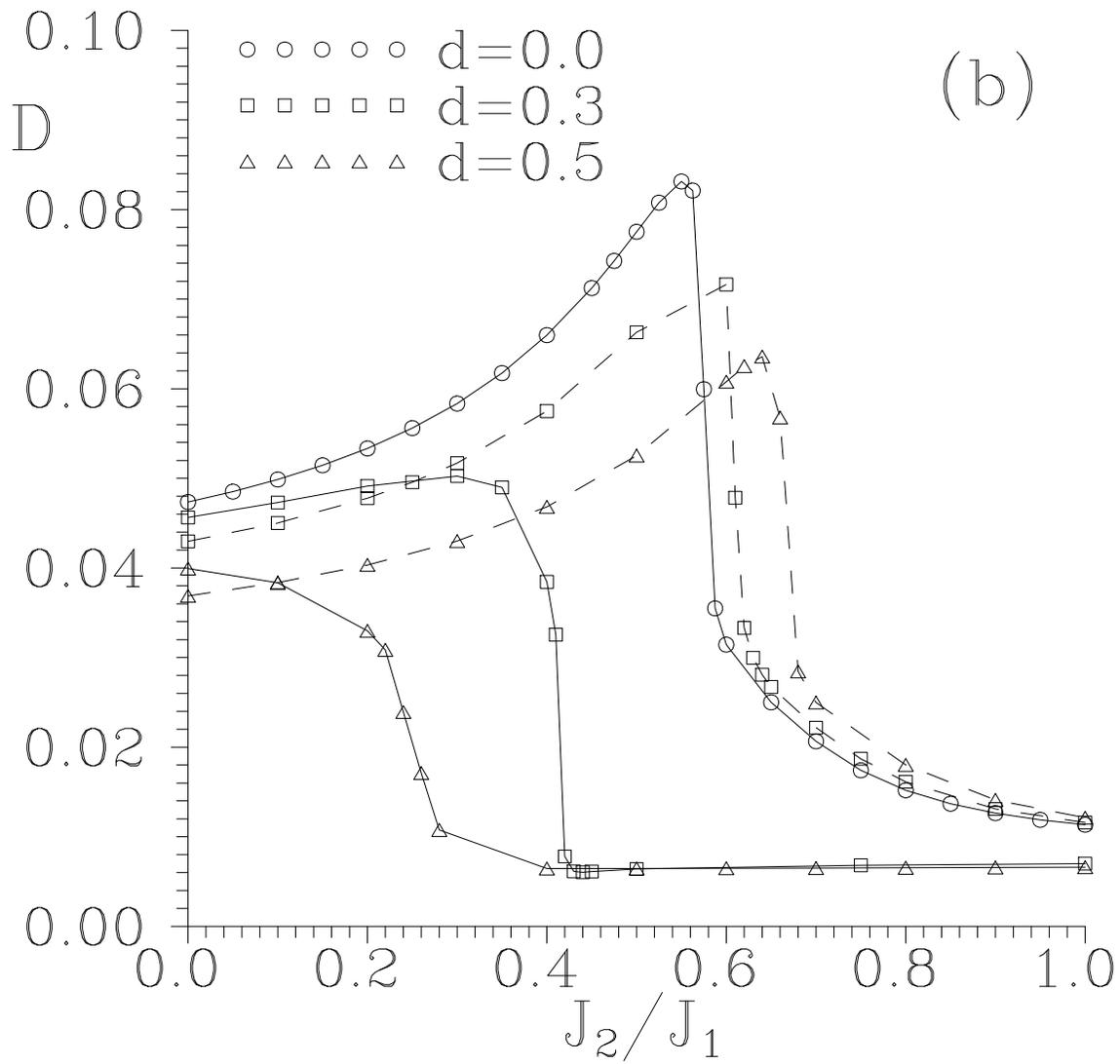
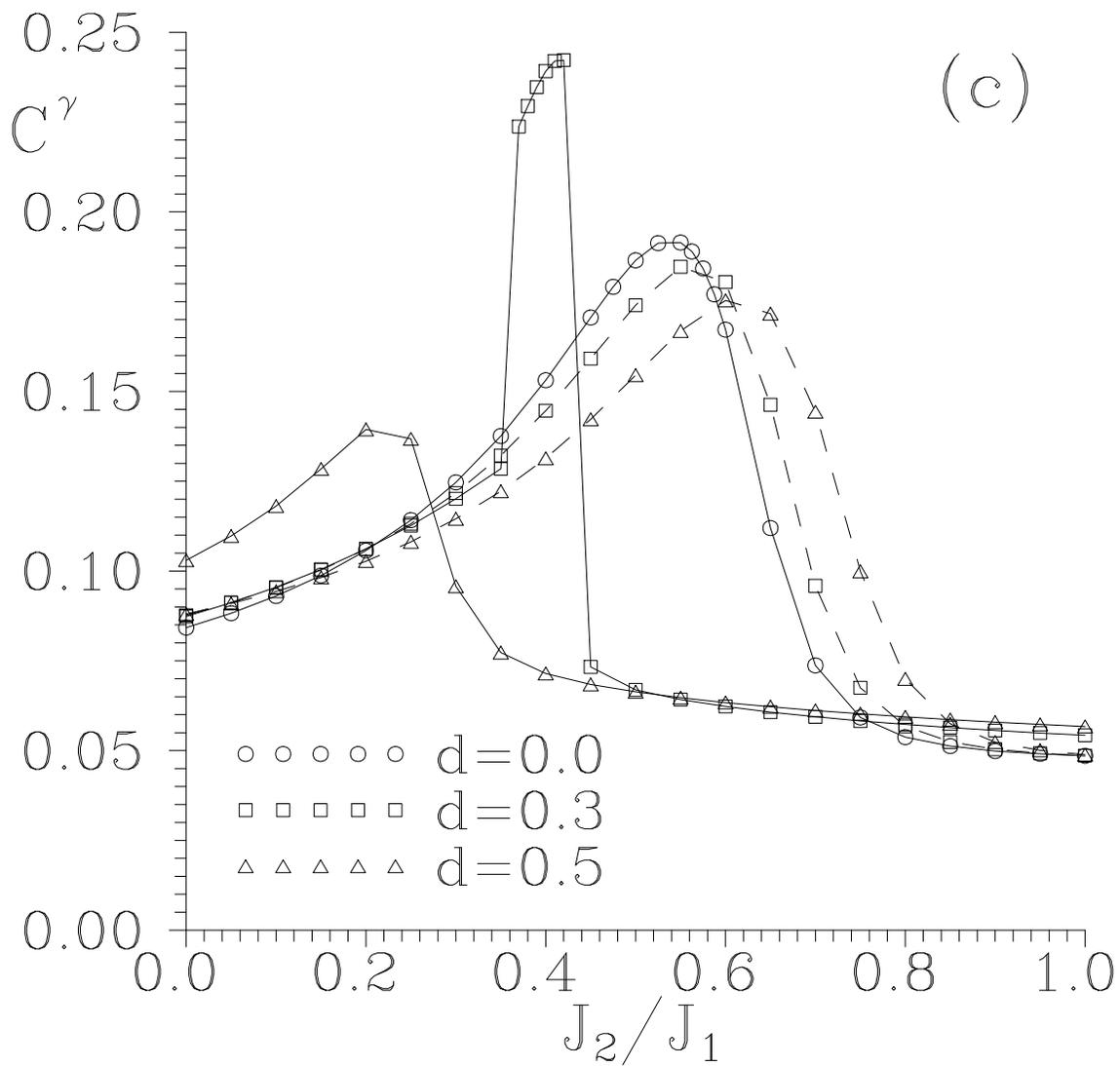
\end{document}